\documentclass[aps,pra,reprint,superscriptaddress]{revtex4-1}
\usepackage{amsmath, amssymb}
\usepackage{graphicx}
\usepackage{xcolor}
\usepackage[caption=false]{subfig}

\newcommand{\ket}[1]{{| {#1}  \rangle}}
\newcommand{\bra}[1]{{\langle {#1} |}}
\newcommand{\expval}[1]{{\langle {#1} \rangle}}

\newcommand{\add}[1]{{#1}}

\begin{document}

\title{Robustness-optimized quantum error correction}

\author{David Layden}
\affiliation{Research Laboratory of Electronics and Department of Nuclear Science and Engineering, Massachusetts Institute of Technology, Cambridge, Massachusetts 02139, USA}

\author{Louisa Ruixue Huang}
\affiliation{Department of Electrical Engineering and Computer Science, Massachusetts Institute of Technology, Cambridge, Massachusetts 02139, USA}

\author{Paola Cappellaro}
\affiliation{Research Laboratory of Electronics and Department of Nuclear Science and Engineering, Massachusetts Institute of Technology, Cambridge, Massachusetts 02139, USA}

\begin{abstract}
Quantum error correction codes are usually designed to correct errors regardless of their physical origins. In large-scale devices, this is an essential feature. In smaller-scale devices, however, the main error sources are often understood, and this knowledge could be exploited for more efficient error correction. Optimizing the quantum error correction protocol is therefore a promising strategy in smaller devices. Typically, this involves tailoring the protocol to a given decoherence channel by solving an appropriate optimization problem. Here we introduce a new optimization-based approach, which maximizes the robustness to faults in the recovery. Our approach is inspired by recent experiments, where such faults have been a significant source of  logical errors. We illustrate this approach with a three-qubit model, and show how near-term experiments could benefit from more robust quantum error correction protocols.
\end{abstract}

\maketitle

\section{Introduction}

The buildup of errors in quantum devices is a central impediment to the development of quantum technologies, such as quantum sensors, networks, and computers. These errors can have a number of different sources, including unwanted coupling to a noisy environment, imperfect controls, and faulty measurements. Quantum error correction (QEC) is a powerful technique for suppressing these various errors. It promises to scale well to large devices in part because it can correct errors without precise knowledge of their physical origins \cite{lidar}. This feature is essential in the long-term, since it would be unfeasible to fully and precisely characterize error mechanisms in large-scale quantum devices. The situation is different in near-term devices, however, where the error mechanisms are often well understood. In these smaller, noisy systems, it could be advantageous to trade the wide net of conventional QEC for a more tailored approach, which exploits knowledge of the dominant error mechanisms to achieve better error suppression \cite{leung:1997, ioffe:2007, cafaro:2014, robertson:2017, layden:2019}.

Optimization-based QEC takes this latter approach \cite{reimpell:2005, yamamoto:2005, fletcher:2007, fletcher:2008, fletcher:2008b, kosut:2009, beny:2010, tyson:2010, taghavi:2010, beny:2011, michael:2016, johnson:2017, albert:2018, li:2019, noh:2019} (see also Ref.~\cite{lidar}, Chapter 13, for a review). It works by mapping the search for good QEC protocols (i.e., codes and recoveries) to an optimization problem, whose solution gives a protocol tailored for a particular type of noise. There are several ways to perform this mapping, some of which enable efficient optimization, as well as a degree of robustness to uncertainties in the error model \cite{lidar, kosut:2008, ballo:2009}. While the resulting protocols often lack an intuitive structure, they hold promise for near-term devices, and perhaps as a first level of encoding in larger devices \cite{fletcher:2008b}.

To date, optimization-based QEC has been largely synonymous with channel-adapted QEC; that is, the focus has been on adapting QEC protocols to the quantum channels describing intrinsic decoherence in idling devices. However, new insights have come from significant experimental advances in implementing QEC since the groundwork for optimization-based QEC was laid. 
A notable feature in some recent, pre-fault-tolerant experiments is that errors due to imperfect QEC recoveries \add{(i.e., measurement and feedback operations)} comprise a significant---if not a limiting---share of the logical errors \cite{ofek:2016, hu:2019}. 
In other words, there is ample room to improve QEC performance in near-term experiments by minimizing the impact of such recovery errors\add{, in the spirit of Refs.\ \cite{rahn:2002, chamberland:2017}}. This suggests a new type of optimization-based QEC, orthogonal to channel-adapted QEC: rather than tailoring QEC protocols to the intrinsic decoherence between recoveries, one could instead find protocols which are optimally robust against imperfections in the recoveries themselves. This is a fundamentally different task; instead of finding an optimal way to suppress errors inherent to a device, it involves devising protocols that perform optimally under imperfect implementation. We demonstrate this latter approach, which we call \textit{robustness-optimized QEC}, by maximizing the robustness of an experimentally-relevant QEC protocol to syndrome measurement errors in the associated recovery.

\section{Setting}

We consider, for illustration, the task of preserving a logical qubit using three physical qubits subject to phase noise, which is the dominant kind of decoherence in many types of quantum devices \cite{biercuk:2009, witzel:2010, bluhm:2011, doherty:2013, muhonen:2014, orgiazzi:2016}. For simplicity, we will not let the QEC code itself vary in the optimization; rather, we will use the phase-flip code, with codewords
\begin{equation}
    \ket{0_\textsc{l}} = \ket{+ \! + \! +}
    \qquad \quad
    \ket{1_\textsc{l}} = \ket{- \! - \! -},
\end{equation}
where $\ket{\pm} = \frac{1}{\sqrt{2}} (\ket{0} \pm \ket{1})$ \add{\cite{kelly:2015, riste:2015, schindler:2011, cramer:2016} (see also Ref.\ \cite{lidar} Ch.~21 and references therein)}. The decoherence can be understood as causing $\sigma_z$ errors on the qubits, which can be detected non-destructively by measuring $\{ P_0, P_1, P_2, P_3 \}$, where $P_0 = \ket{0_\textsc{l}} \! \bra{0_\textsc{l}} + \ket{1_\textsc{l}} \! \bra{1_\textsc{l}}$ and $P_j = Z_j P_0 Z_j$ are rank-2 orthogonal projectors. ($Z_j$ denotes the Pauli matrix $\sigma_z$ on qubit $j$.) A $Z_j$ error will transform the logical state $\ket{\psi_\textsc{l}} = \alpha \ket{0_\textsc{l}} + \beta \ket{1_\textsc{l}}$ into range$(P_j)$ in a way that can be reversed by applying $Z_j$. The quantum channel describing this ideal recovery procedure is
\begin{equation}
    \mathcal{R}_\text{ideal}(\rho)
    =
    \sum_{j=0}^3 U_j^\dagger P_j \rho P_j U_j,
    \label{eq:R_ideal}
\end{equation}
where $U_0 = I$, $U_j = Z_j$ for $j \ge 1$ \cite{nielsen}.  \add{Note that throughout this work we consider the conceptually-simple QEC strategy in which errors are physically corrected upon detection, as opposed to more sophisticated strategies using Pauli/Clifford frames \cite{divincenzo:2007, chamberland:2018}.}

Suppose, however, that the measurement process is imperfect, and reports the wrong result uniformly with some  probability $p_\text{meas}$, e.g., due to an error on an uncorrected ancilla. That is, a general state may be projected into range($P_j$) in the usual way, but the measurement device sometimes reports it to be in range($P_k$) for $k\neq j$. Feeding back on this faulty syndrome would cause a logical error. The channel describing this imperfect recovery is \footnote{Note that a syndrome measurement error is not equivalent to a $Z_j$ error on a data qubit, since it has no effect in the absence of feedback.}:
\begin{equation}
    \mathcal{R}_\text{faulty}(\rho)
    =
    (1-p_\text{meas}) \, \mathcal{R}_\text{ideal} (\rho)
    +
    \frac{p_\text{meas}}{3} \sum_{\substack{i,j=0 \\ i \neq j }}^3
    U_j^\dagger P_i \rho P_i U_j.
    \label{eq:R_faulty}
\end{equation}
\add{Note that $p_\text{meas}$ is the total measurement error probability, which may encompass the individual error probabilities from measurements on several ancilla qubits.}

How can the phase-flip code be made more robust to such imperfections in the recovery? One can imagine two extreme strategies which work well in different regimes:
\begin{description}
\item[Strategy A - Conventional QEC] If $p_\text{meas}$ is sufficiently small, a good strategy is to periodically perform $\mathcal{R}_\text{faulty}$, and simply accept the performance degradation due to non-zero $p_\text{meas}$.

\item[Strategy B - Quantum Zeno Effect] If $p_\text{meas}$ is sufficiently large, it may be better not to actively correct phase errors at all. Instead, one could suppress them---independent of $p_\text{meas}$---through the quantum Zeno effect by repeatedly measuring $\{ P_j \}$ without feedback \add{\cite{degasperis:1974, misra:1977, vaidman:1996, erez:2004}}.
\end{description}
Which of these represents the better approach will depend both on $p_\text{meas}$ and on the total amount of time, $\Delta t$, for which one wants to preserve the logical state.

More generally, however, one could interpolate between Strategies A and B as follows: with probability $p_\text{fb}$ perform $\mathcal{R}_\text{faulty}$, and with probability $1-p_\text{fb}$ measure the parity $\{P_j \}$ but do not feed back. This corresponds to the channel
\begin{equation}
    \mathcal{R}_\text{opt}(\rho) = p_\text{fb} \, \mathcal{R}_\text{faulty}(\rho)
    +
    (1-p_\text{fb}) \sum_{j=0}^3 P_j \rho P_j.
\end{equation}
Strategies A and B then correspond to $p_\text{fb}=1$ and $0$ respectively. Instead of adopting either strategy entirely, we will treat $p_\text{fb}$ as a free parameter, and find the optimal value which maximizes robustness to recovery imperfections. For certain values of $p_\text{meas}$ and $\Delta t$, we find that intermediate values of $p_\text{fb}$ outperform both extreme strategies.

\section{Decoherence Model and Objective Function}

A common and simple model for the phase noise is a Lindblad equation with $Z_j$ jumps. This would be equivalent to the qubits' energy gaps being subject to a zero-mean Gaussian white noise process, and would suppress single-qubit coherence as $ \big| \bra{0} \rho_j \ket{1} \big| \propto \exp(-t/T_2^*)$ for some characteristic dephasing time $T_2^*$ \cite{kubo:1962, layden:2018}. While this is a common idealization of realistic decoherence, it is unsuitable here. The quantum Zeno effect---which has been observed in several experiments, including some which preserve subspaces of dimension $\ge 1$, see e.g., \cite{itano:1990, fischer:2001, bernu:2008, schafer:2014, signoles:2014}---does not occur in the pathological limit where the phase noise has infinite power at high frequencies. This is precisely the limit described by the aforementioned Lindblad model, and so repeated measurements of $\{ P_j \}$, no matter how frequent, would not preserve a logical state in this model. Adopting such a model would make it largely pointless to optimize $p_\text{fb}.$

A more realistic model for some experiments, which displays a Zeno effect and in turn a rich landscape in $p_\text{fb}$, is dephasing due to low-frequency noise in the qubits' energy gaps. Such noise suppresses single-qubit coherence as $\exp[ -(t/T_2^*)^2]$, which is more typical in many experiments with slowly-evolving environments \cite{ramsey, suter:2016, degen:2017}. Concretely, we assume that in a suitable frame the qubits evolve as
\begin{equation}
    H(t) = \frac{1}{2} \sum_{j=1}^3 \omega_j(t) Z_j,
\end{equation}
where the $\omega_j$'s are independent quasi-static noise processes that are approximately constant over $[0, \Delta t]$ but vary between runs of the experiment. More precisely, we take $\omega_j$ to be a zero-mean, stationary Gaussian stochastic process with a constant autocorrelation function
\begin{equation}
\expval{\omega_j(t) \, \omega_j(0)} = \frac{2}{(T_2^*)^2},
\end{equation}
where $\expval{ \cdot }$ denotes a (classical) average over realizations of $\omega_j$. That is, the power spectrum of $\omega_j$ goes as $S_{\omega_j}(\nu) \propto \delta(\nu)$. While the dynamics in each run of the experiment is unitary, the average dynamics is not, which leads to dephasing. \add{Note that dynamical decoupling would be useful in refocusing this noise, although we will not consider it here in order to isolate the effects of QEC \cite{viola:1998,  ban:1998,  biercuk:2011}. In practice, however, it could be beneficial to use dynamical decoupling in conjunction with the present QEC scheme.}

We suppose that one can perform $\mathcal{R}_\text{opt}$ $n \ge 1$ times, equally spaced, during the interval $[0,\Delta t]$ (with the first $\mathcal{R}_\text{opt}$ occurring at time $\Delta t/n$ and the last at $\Delta t$). To describe the effect of this procedure, we first define the superoperator $\mathcal{V}_t(\rho) := V_t \, \rho \, V_t^\dagger$, where 
\begin{equation}
V_t := \exp \left[-i \int_0^t H(t') \, dt' \right].
\end{equation}
Then, if the system is prepared in the initial logical state $\rho_\textsc{l} = \ket{\psi_\textsc{l}} \! \bra{\psi_\textsc{l}}$, its final state after performing $n$ repetitions of $\mathcal{R}_\text{opt}$ in the interval $[0, \Delta t]$ is
\begin{equation}
    \rho_f = \left \langle \, 
    \left( \mathcal{R}_\text{opt} \, \mathcal{V}_{\Delta t/n} \right)^n
    \, \right \rangle (\rho_\textsc{l}).
\end{equation}
We will use the quantum fidelity $F = \bra{\psi_\textsc{l}} \rho_f \ket{\psi_\textsc{l}}$ as a measure of performance. More precisely, we use the fidelity averaged over all initial logical states, $\overline{F}$, as a figure of merit/objective function when optimizing the robustness. For $n=1$ recovery (at a final time $\Delta t$), we have
\begin{align}
    \overline{F}_1 &=
    \frac{1}{6} \big[1+p_\text{fb}(3-4 p_\text{meas}) \big]  +
    \frac{1}{2} e^{-2 (\Delta t/T_2^*)^2} (1-p_\text{fb}) \nonumber \\
    &+
    \frac{1}{4} e^{-(\Delta t/T_2^*)^2} \big[1+p_\text{fb}(1-2p_\text{meas}) \big] \\
    &+
    \frac{1}{12}e^{-3 (\Delta t/T_2^*)^2} \big[1+p_\text{fb}(2p_\text{meas} - 3) \big]. \nonumber
\end{align}
We were able to find analytic expressions for $\overline{F}_n$ with $1 \le n \le 10$, although  for $n \ge 2$ the expressions quickly become lengthy and so have been relegated to the supplementary material \footnote{The expression for $\overline{F}_{10}$, for instance, contains 4588 terms. It, along with the other $\overline{F}_n$'s, can be found in the Mathematica notebook included in the supplementary material.}. Average fidelities for $n \ge 11$ are not only difficult to compute, but they are of limited relevance to near-term experiments where control limitations and other sources of error impose a limit on $n$. Moreover, even in the longer term, the number of recoveries within an interval  $[0,\Delta t]$ must be limited if there is to be time left over to perform logical operations on the encoded state (since recoveries will not be instantaneous in practice).

\section{Results}

\begin{figure}[h!]
    \centering
    \includegraphics[width=0.46\textwidth]{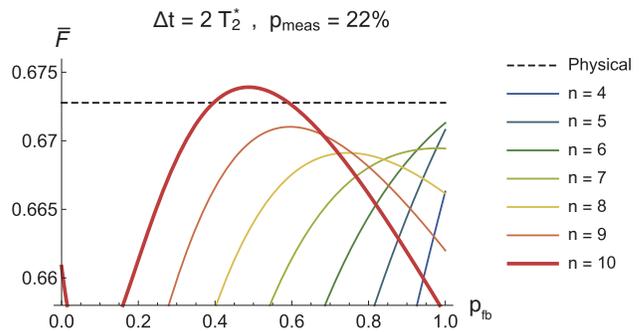}
    \caption{The average fidelity versus $p_\text{fb}$ and $n$ for $\Delta t = 2 T_2^*$ and $p_\text{meas}=0.22$. The solid lines denote $\overline{F}_n$ for $n\ge 4$; the curves for $n \le 3$ are not visible as they are too low. The dashed line is the fidelity of single physical qubit under the same noise. The optimal strategy of those considered, that is, the $n \in [1,10]$ and $p_\text{fb} \in [0,1]$ combination producing the highest fidelity, uses $n=10$ (bold red line) and $p_\text{fb} = 0.488$ to achieve a fidelity of $\overline{F}_\text{max} = 0.674$.}
    \label{fig:hybrid}
\end{figure}

We will treat $\Delta t$ and $p_\text{meas}$ as fixed in any given experiment, which leaves the parameters $n$ and $p_\text{fb}$ to be optimized. The dependence of $\overline{F}_n$ on these parameters, for a particular $\Delta t$ and $p_\text{meas}$, is illustrated in Fig.~\ref{fig:hybrid}. For this $\Delta t$ and $p_\text{meas}$, the most robust strategy is a hybrid of Strategies A and B, which outperforms the two extremes. Perhaps counter-intuitively, this means that the average fidelity is \textit{increased} here by introducing extra randomness into $\mathcal{R}_\text{opt}$ through the choice of $0 < p_\text{fb} <1$.

\begin{figure}[h!]
    \centering
    \includegraphics[width=0.46\textwidth]{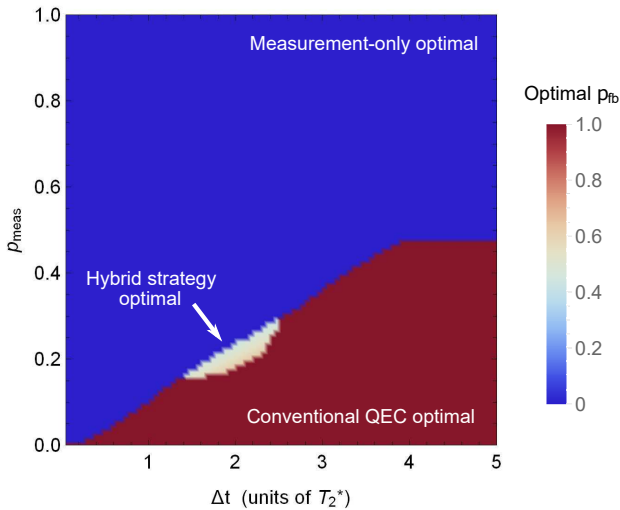}
    \caption{The optimal $p_\text{fb}$ for different values of $\Delta t$ and $p_\text{meas}$, \add{after encoding}. The best $p_\text{fb}^{(n)}$ for each $\overline{F}_n$ was found separately; this figure shows the one giving the highest value of $\overline{F}_n$. $p_\text{fb}=1$ gives the conventional QEC strategy of measurement and feedback, whereas $p_\text{meas}=0$ uses no feedback, relying instead on a quantum Zeno effect from repeated parity measurements.}
    \label{fig:pfb_opt}
\end{figure}

More generally, for each $(\Delta t, p_\text{meas})$, we optimize $\overline{F}_n$ over both $n$ and $p_\text{fb}$. The optimal $p_\text{fb}$, shown in Fig.~\ref{fig:pfb_opt}, has three distinct ``phases" in the parameter range considered. As anticipated above, when $p_\text{meas}$ is sufficiently small the optimal strategy is to perform conventional recoveries ($p_\text{fb}=1$) and simply accept the occasional faults that these introduce. Conversely, when $p_\text{meas}$ is sufficiently large (and/or $\Delta t$ is sufficiently small), it is better to avoid feedback entirely and simply preserve the logical state using a Zeno effect from repeated parity measurements. We observe a sharp transition between these two optimal strategies in much of the parameter space. Mathematically, this is due to the maxima of $\overline{F}_n$ often occurring on the boundary of $\{ p_\text{fb} \in [0,1] \}$ rather than in the interior. Remarkably, however, there is a finite region where the transition is not sharp, which exhibits a third ``phase" corresponding to optimal $p_\text{fb}$'s near 0.5 (though not always exactly equal to 0.5, see e.g., Fig.~\ref{fig:hybrid}). The $\Delta t$ and $p_\text{meas}$ from Fig.~\ref{fig:hybrid} are from this region.

\begin{figure*}
    \centering
    \subfloat{
    \includegraphics[width=0.335\textwidth]{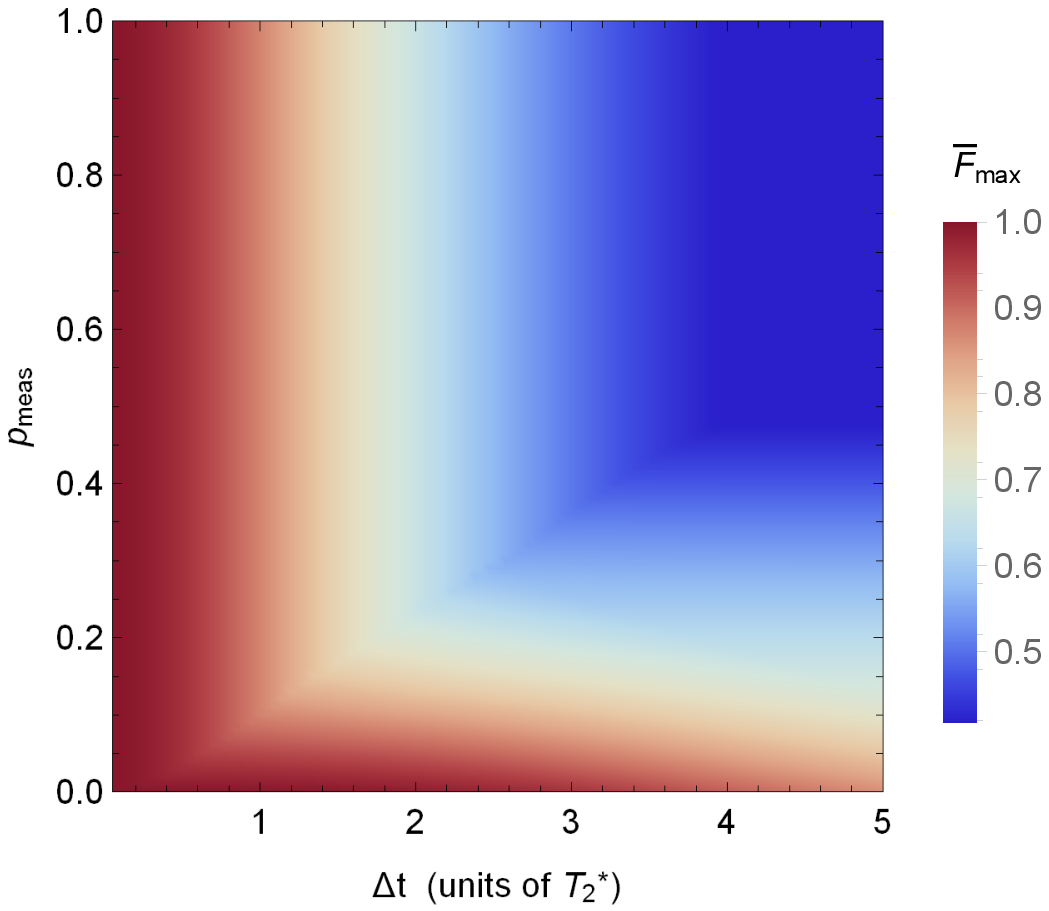}
    }
    \subfloat{
    \includegraphics[width=0.35\textwidth]{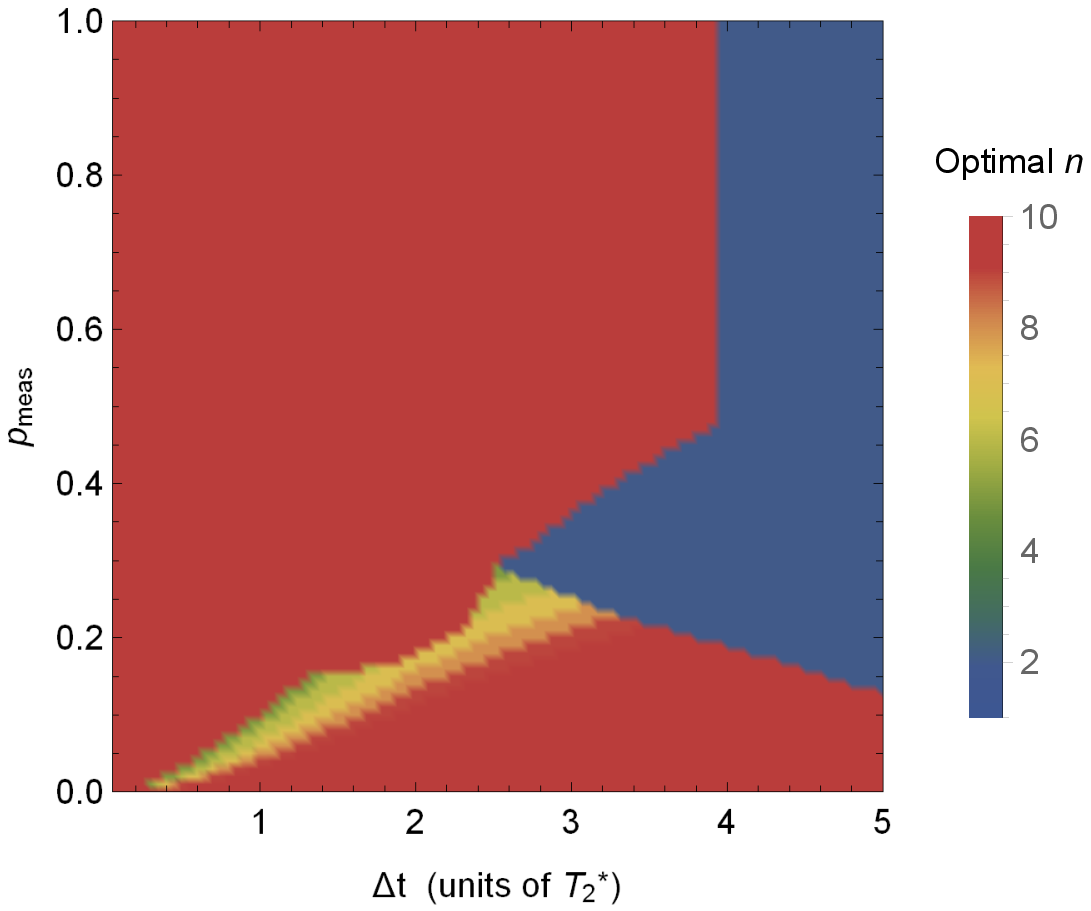}
    }
    \subfloat{
    \raisebox{2.5em}{
    \includegraphics[width=0.27\textwidth]{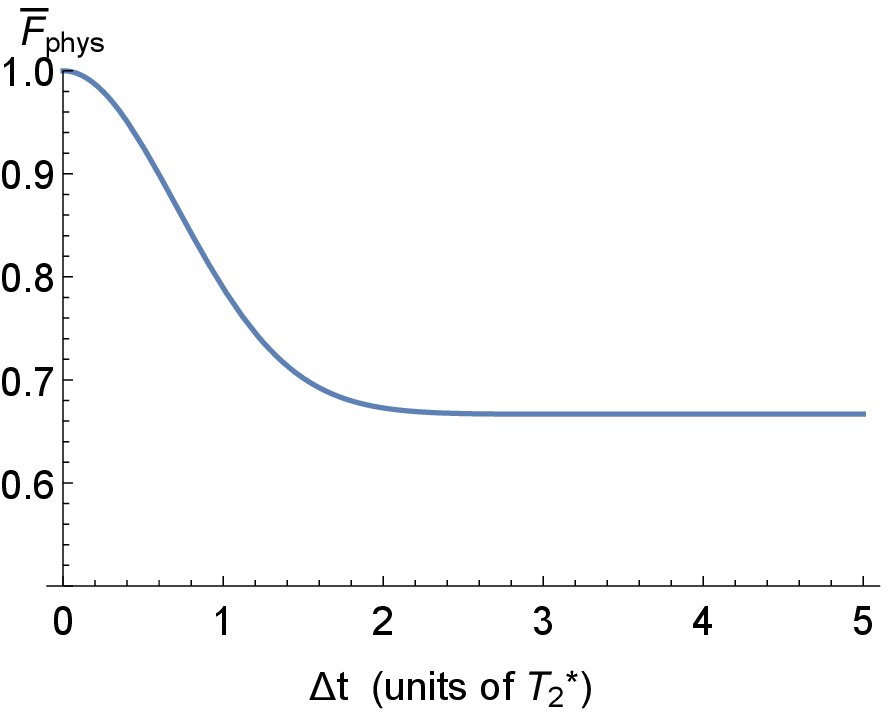}
    }}
    \caption{Left: The maximum fidelity achievable by optimizing over $p_\text{fb} \in [0,1]$ and $1 \le n \le 10$. Center: The optimal $n$ which gives this maximum fidelity. Right: For comparison, the fidelity for a single physical qubit subject to the same noise.}
    \label{fig:F_and_n}
\end{figure*}

The maximum values of $\overline{F}_n$ and the optimal $n$'s resulting from this same optimization are shown in the left and center panels of Fig.~\ref{fig:F_and_n}. As one might expect, the fidelity decays gradually with increasing $\Delta t$ and $p_\text{meas}$. The choice of $n$ is more complex, as the same optimal $n$ can represent different strategies depending on the corresponding $p_\text{fb}$. For instance, using a large $n$ is optimal both when $p_\text{meas}$ is small and when it is large (compared to $\Delta t$). 
In the former regime one has $p_\text{fb}=1$, so a large $n$ reduces the buildup of uncorrectable errors of weight 2 and 3 due to phase noise. In the latter regime $p_\text{fb}=0$, so a large $n$ means frequent measurements and therefore a stronger Zeno effect. Between these two regimes, moderate values of $n$ are optimal, as they provide some correction without too many recovery faults. Finally, for large $\Delta t$ and large $p_\text{meas}$ we find small $n$ to be optimal. This is likely an artifact of considering only $n \le 10$: $\lim_{n\rightarrow \infty} \overline{F}_n = 1$ for all $\Delta t$ and $p_\text{meas}$, so if we allowed unbounded $n$ the Zeno strategy would always be optimal in principle. 
However, for large $\Delta t$, $n\le 10$ measurements are insufficient to produce a strong Zeno effect, so the next-best strategy is to use faulty recoveries sparingly. \add{Note finally that for large $\Delta t$ and/or $p_\text{meas}$, including some values where a hybrid strategy is shown to be optimal in Fig.~\ref{fig:pfb_opt}, it may be better not to perform encoding at all (see appendix).}

\section{Conclusions and Outlook}

We have shown that one can optimize the robustness of small, \add{pre-fault-tolerant }QEC protocols to recovery errors, in analogy to how such protocols have previously been optimized for specific decoherence channels. Whereas the latter approach is often called channel-adapted QEC, we term ours robustness-optimized QEC. Errors from QEC recoveries have formed a significant fraction of the total logical errors in recent experiments \cite{ofek:2016, hu:2019}. This suggests that there is much to be gained by optimizing for robustness against such errors instead of---or as well as---optimizing for the decoherence inherent in particular devices. \add{While fault-tolerant methods could handle such errors in the longer term, the present strategy is specifically intended for nearer-term, pre-fault-tolerant experiments \cite{fowler:2012, campbell:2019}.}

These results raise a number of further questions and possibilities, which we divide into technical points and points of strategy. First the technical points. As in previous works on optimization-based QEC, there is some ambiguity here in choosing a figure of merit. We have used average fidelity for convenience; however, the optimization could give slightly different results/strategies if we had chosen a different objective function, e.g., trace distance to the identity \add{\cite{iyer:2018}}. Moreover, there is \add{often} little reason to favor one particular performance measure over another \textit{a priori} (see \cite{lidar}, Chapter 13). It would be useful to better understand \add{how such effects affect schemes of the sort considered here}. Similarly, the robust QEC strategies found here are robust against a particular type of error during recovery, which we chose as a generic illustration---they are not a panacea \footnote{In particular, our fault model is different---and simpler---than the dominant recovery imperfections in \cite{ofek:2016, hu:2019}.}. Different types of recovery errors will likely require different models and optimization mappings than the ones used here, which may need to be worked out case-by-case. Fortunately, there is less ambiguity with this choice, since the dominant error sources in current experiments are often well-understood (see, e.g., \cite{ofek:2016, hu:2019}). There is likely more room for optimization in more detailed fault models, e.g., where the probability of measurement errors is outcome-dependent\add{, or when such errors are predominantly due to decoherence of ancillas (rather than limited measurement fidelity, for instance)} \cite{cramer:2016, kelly:2015}. \add{Indeed, noise that is highly structured can often be dealt with more efficiently in general \cite{biercuk:2011, tuckett:2018, tuckett:2019, leung:1997,ioffe:2007,cafaro:2014, layden:2019}.} Finally, previous works on channel-adapted QEC have introduced sophisticated mappings which result in convex/bi-convex optimization problems that are efficiently solvable. Developing analogous tools for robustness-optimized QEC would enable the analysis of more complex codes and even more realistic noise models (such as $1/f$ noise) than those analyzed here (see \cite{fletcher_thesis} and references therein).

As for the points of strategy: First, rather than optimizing the probability of performing feedback, one could instead optimize over deterministic strategies of the form ``feedback, no feedback, feedback, $\dots$". This would most likely improve performance, but at the cost of transforming a continuous optimization problem into a potentially more expensive combinatorial one. Second, while we have only optimized the form of the recovery here, it may be advantageous to optimize both the code and the recovery, as is common in channel-adapted QEC \cite{fletcher_thesis}. Moreover, one could think of changing the recovery's structure more generally, e.g., by using different $U_j$'s in Eqs.~\eqref{eq:R_ideal} and \eqref{eq:R_faulty}. (However, we have had limited success with this approach to date.) Finally, it may be possible to build upon the existing machinery of channel-adapted QEC by incorporating tools from robust or stochastic optimization, which can find near-optimal solutions to problems that are robust against imperfections in implementation \cite{ben:2009} (see also \cite{teo_thesis} for an introduction). There appears to be ample room for new approaches to optimization-based QEC in light of recent experimental progress.

\section{Acknowledgements}

We wish to thank Steven Girvin, Liang Jiang and Stefan Krastanov for helpful discussions. This work was supported in part by NSF grants EFRI-ACQUIRE 1641064 and CUA PHY1734011.

\bibliography{references}

\begin{thebibliography}{68}
\bibitem [{ram()}]{ramsey}%
  \BibitemOpen
  \href@noop {} {}\bibinfo {note} {This is the same Ramsey decay as produced by
  $1/f$-type noise, where most of the noise power is at low
  frequencies.}\BibitemShut {Stop}%
\end{thebibliography}

\appendix
\section{Optimization results for each $\boldsymbol{n}$}

\begin{figure*}
    \subfloat{
    \includegraphics[height=0.4\textwidth]{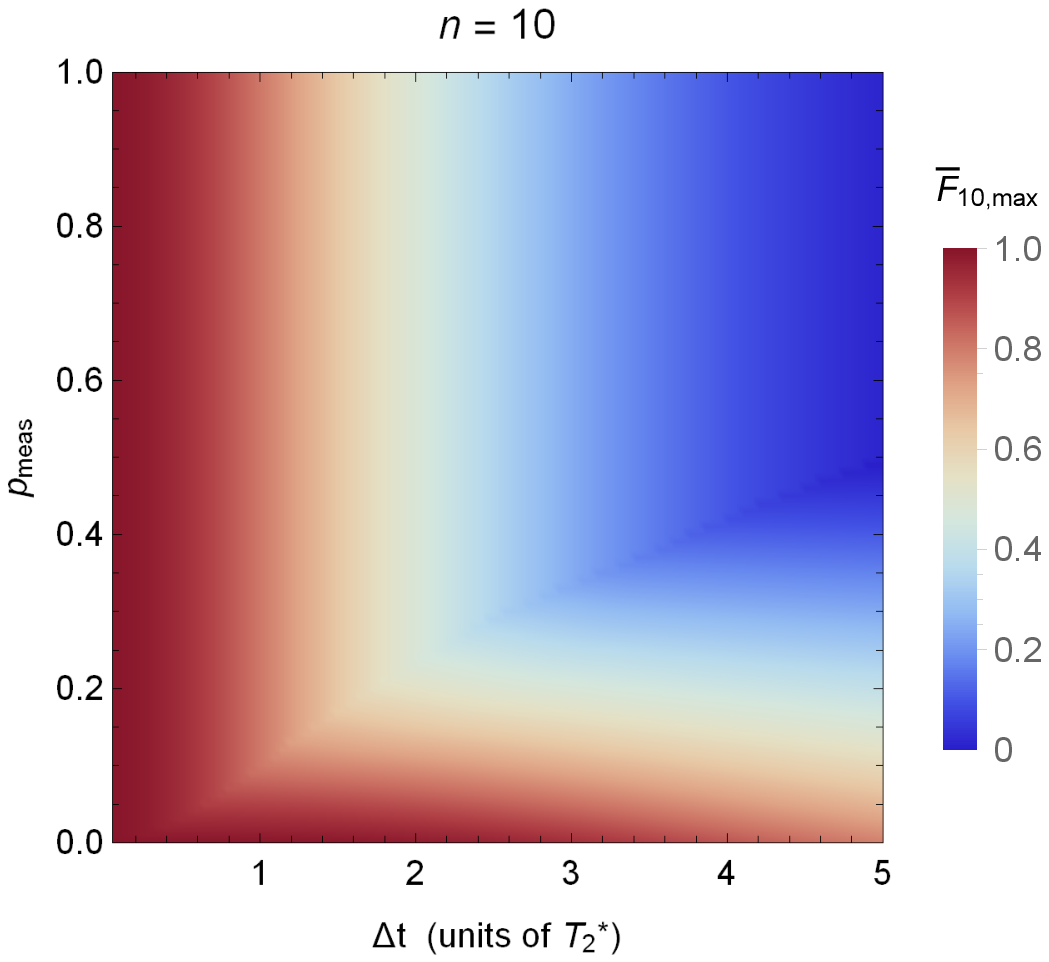}
    }
    \hfill
    \subfloat{
    \includegraphics[height=0.4\textwidth]{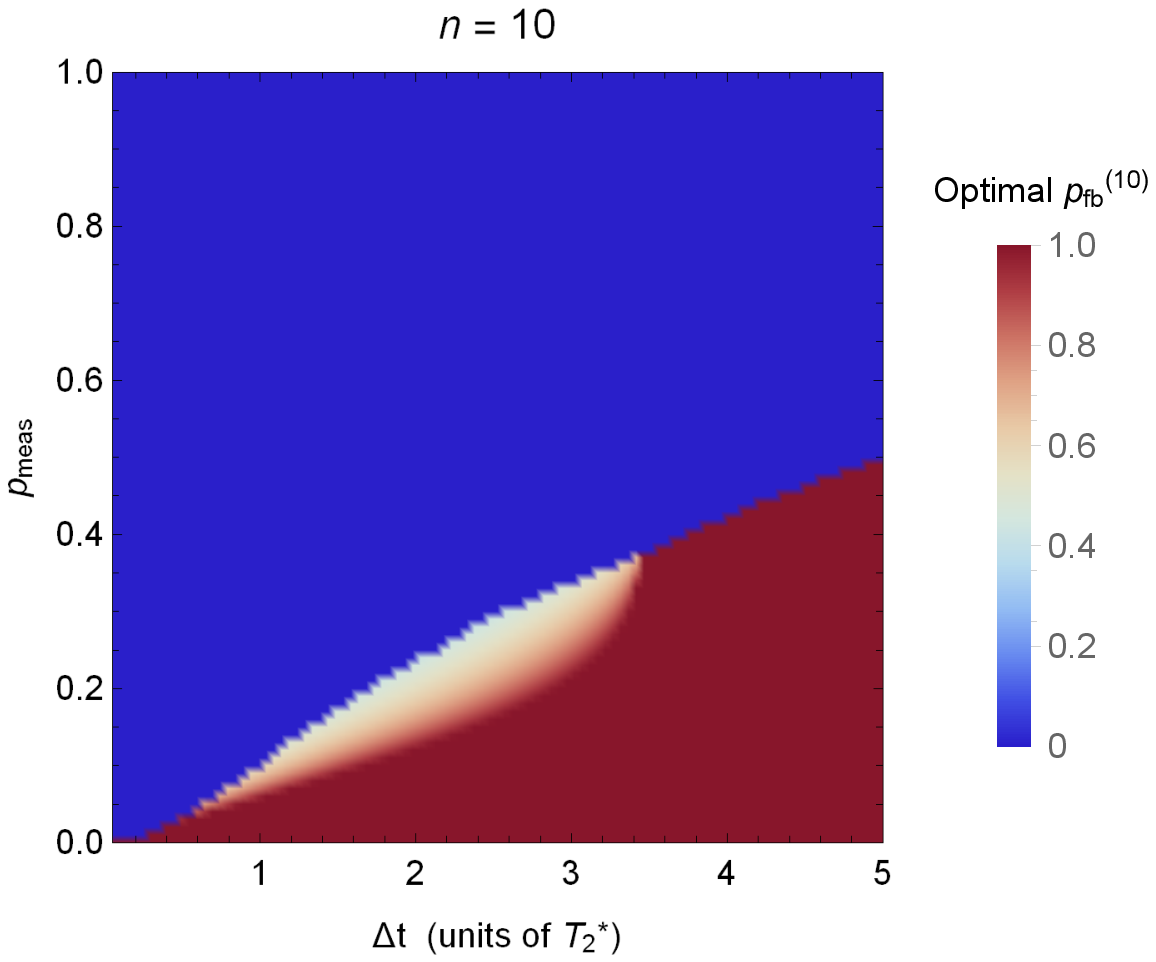}
    }
    \caption{The optimal $\overline{F}_n$ for each $1 \le n \le 10$ separately (left panels), and the corresponding $p_\text{fb}^{(n)}$ at which this fidelity is achieved (right panels). Note that the color bars in the left panels have a different scale than that in Fig.~\ref{fig:F_and_n}.}
    \label{fig:individual}
\end{figure*}

Figs.~\ref{fig:pfb_opt} and \ref{fig:F_and_n} show the results of an optimization performed first over $p_\text{fb} \in [0,1]$ for each $n$, and then over $1 \le n \le 10$. In Fig.~\ref{fig:individual} we show the results from the first step of this optimization separately for each $n$.

\begin{figure*}
    \centering
    \subfloat{
    \includegraphics[height=0.4\textwidth]{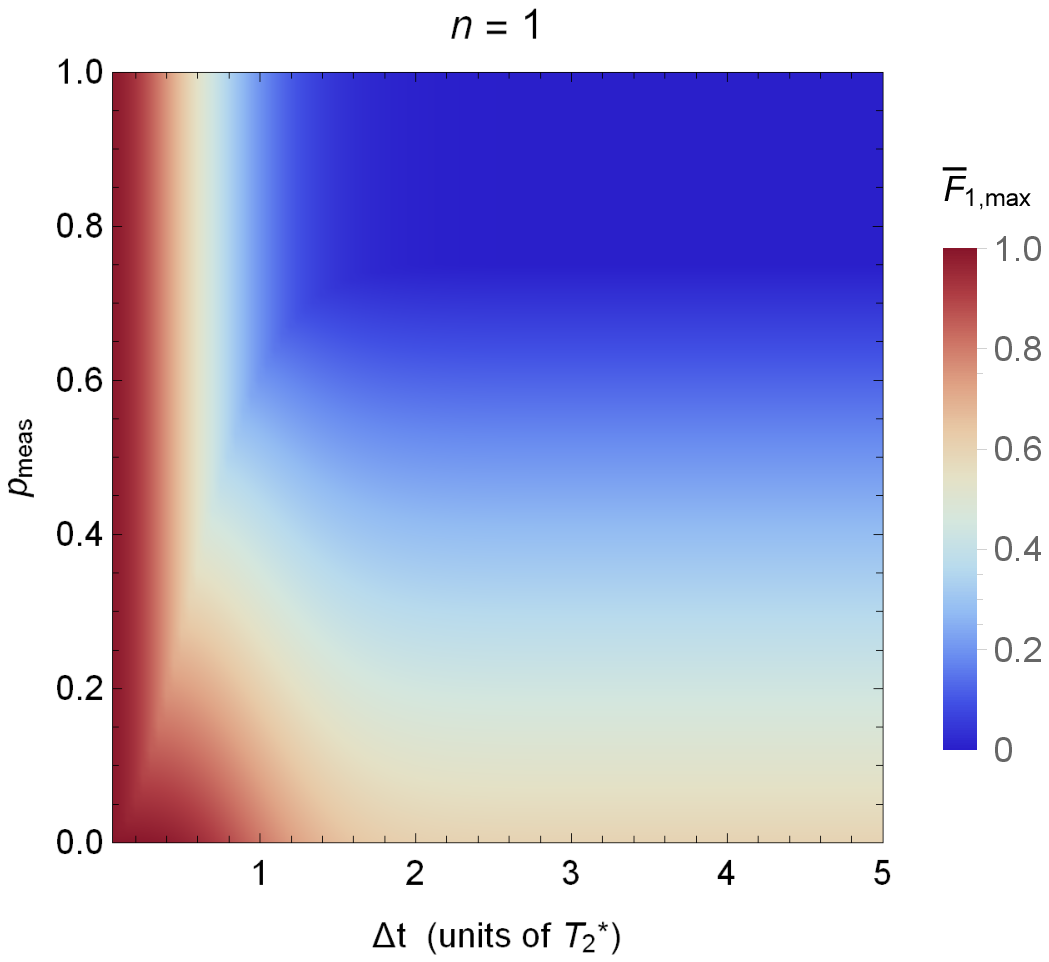}
    }
    \hfill
    \subfloat{
    \includegraphics[height=0.4\textwidth]{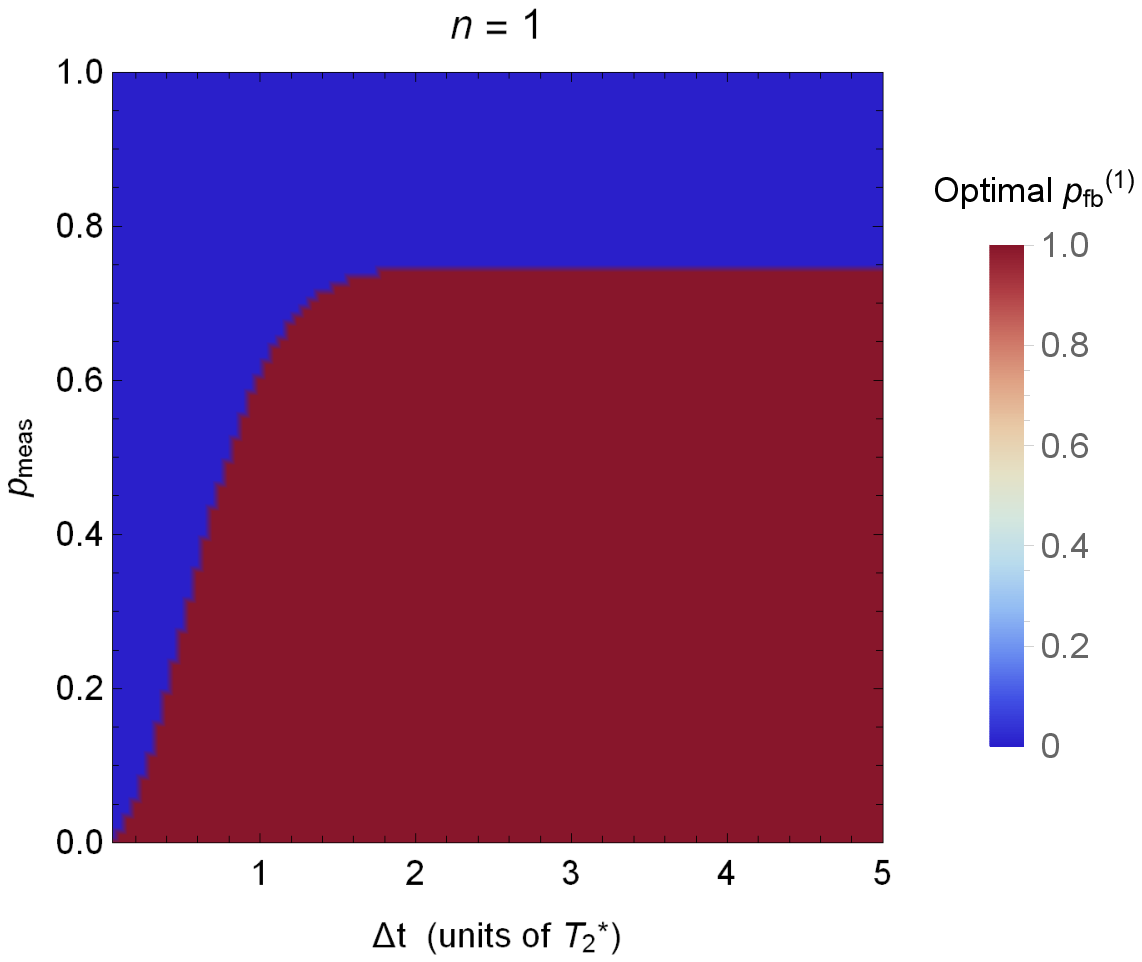}
    }
\end{figure*}
\begin{figure*}
    \subfloat{
    \includegraphics[height=0.4\textwidth]{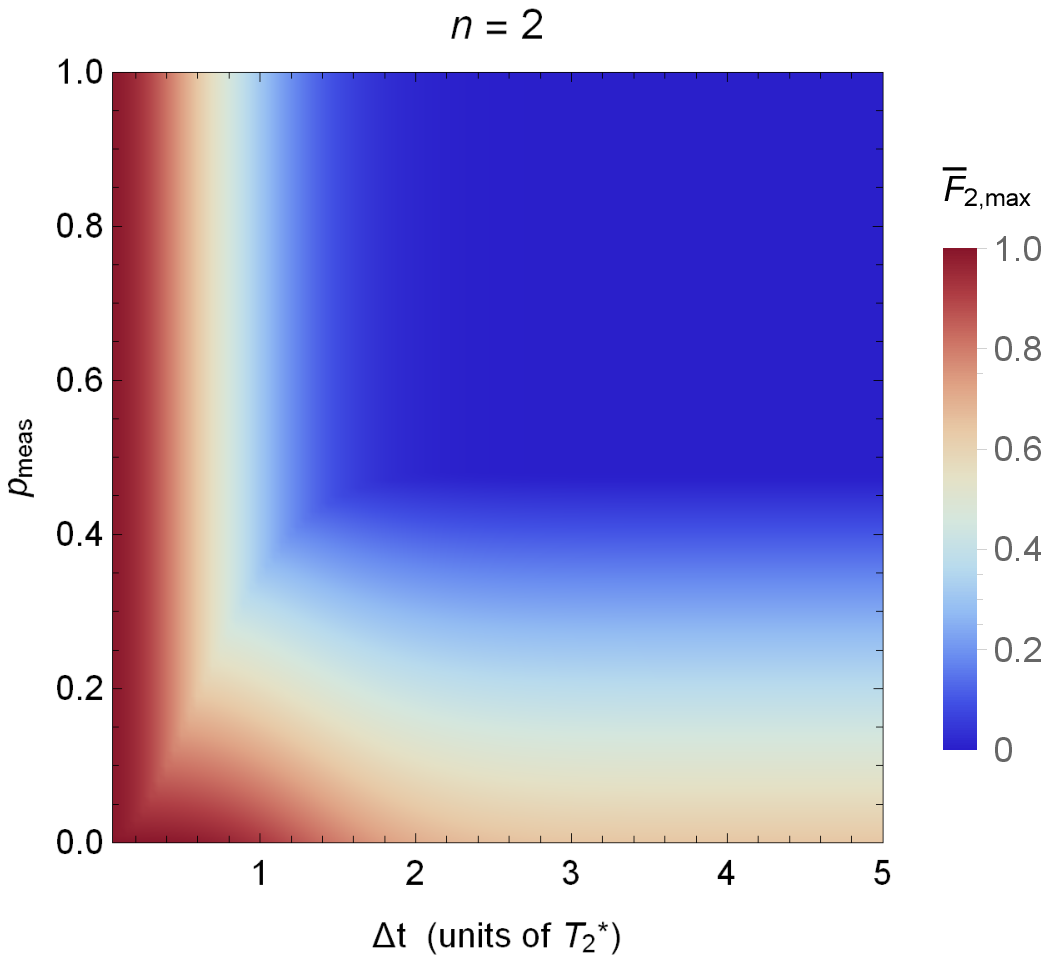}
    }
    \hfill
    \subfloat{
    \includegraphics[height=0.4\textwidth]{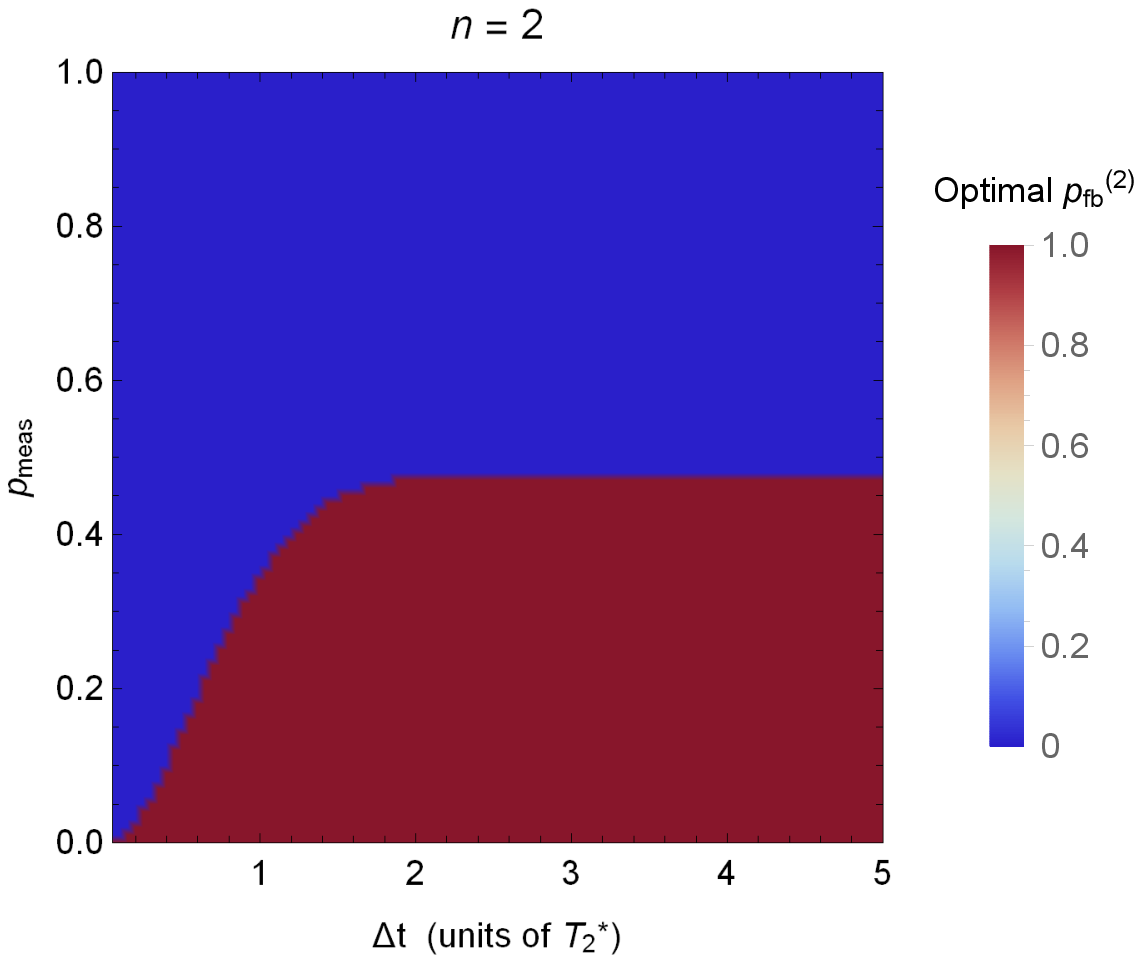}
    }
\end{figure*}
\begin{figure*}
    \subfloat{
    \includegraphics[height=0.4\textwidth]{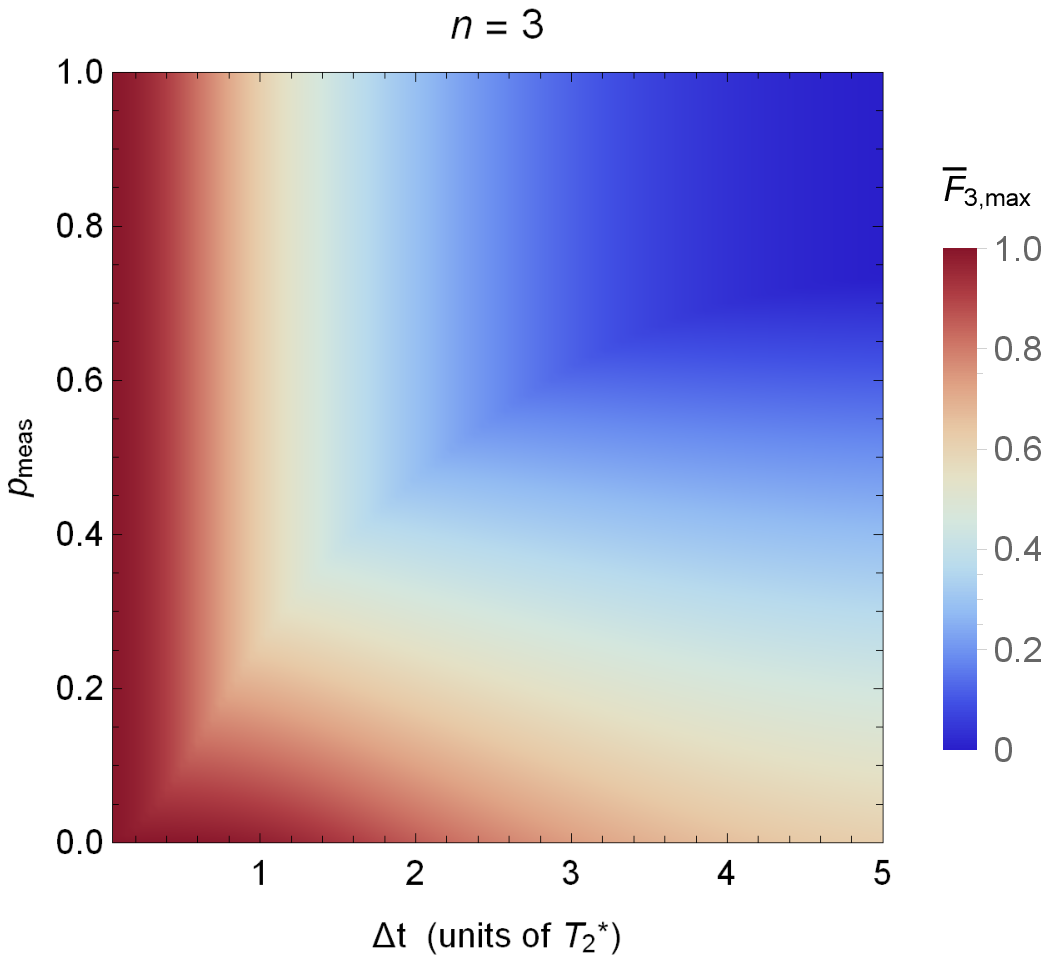}
    }
    \hfill
    \subfloat{
    \includegraphics[height=0.4\textwidth]{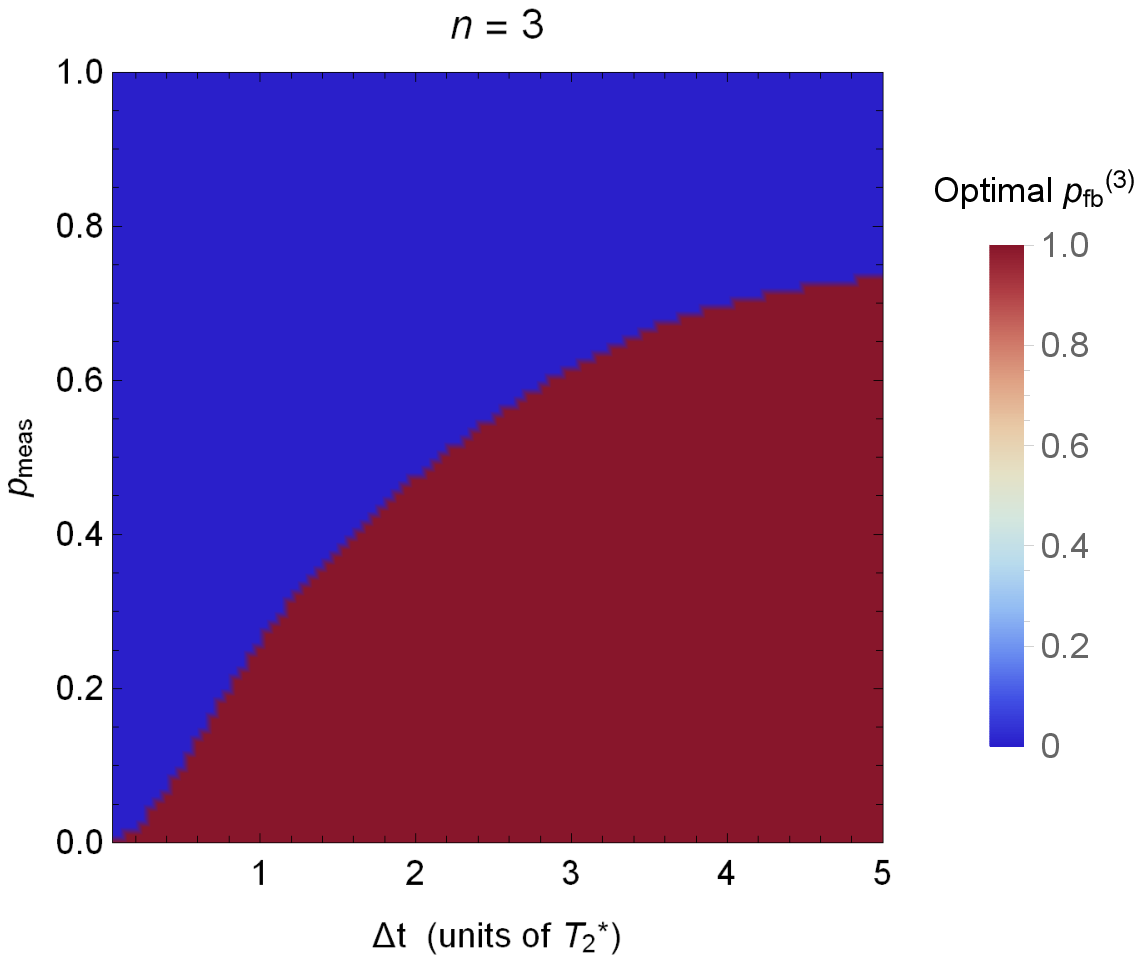}
    }
\end{figure*}
\begin{figure*}
    \subfloat{
    \includegraphics[height=0.4\textwidth]{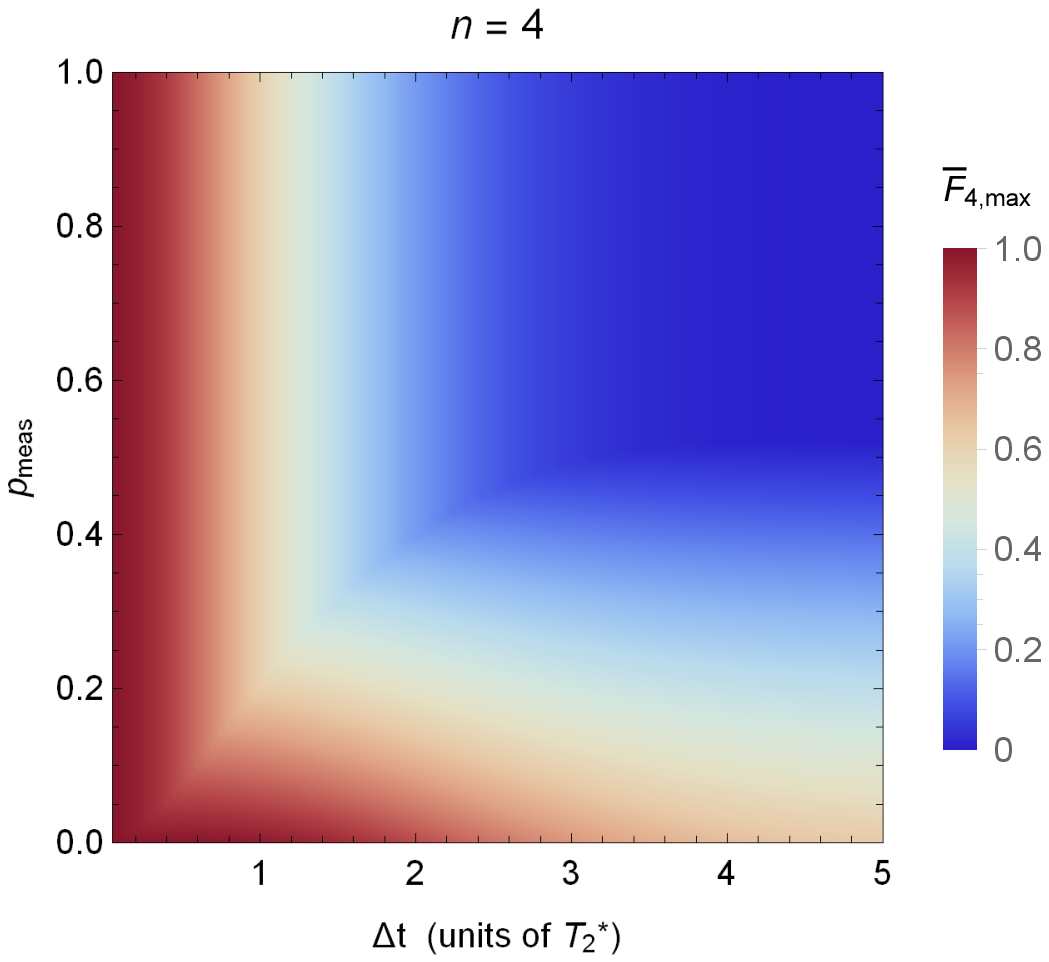}
    }
    \hfill
    \subfloat{
    \includegraphics[height=0.4\textwidth]{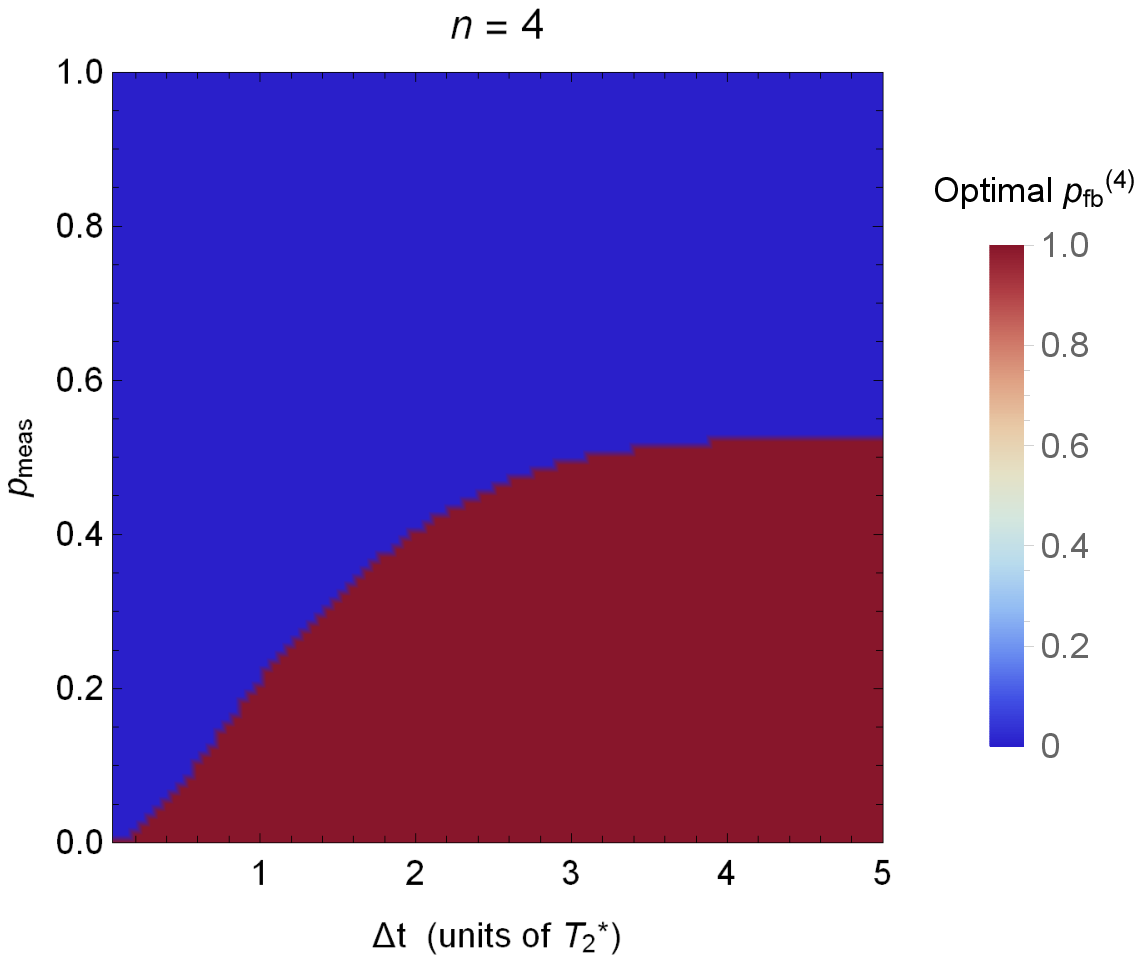}
    }
\end{figure*}
\begin{figure*}
    \subfloat{
    \includegraphics[height=0.4\textwidth]{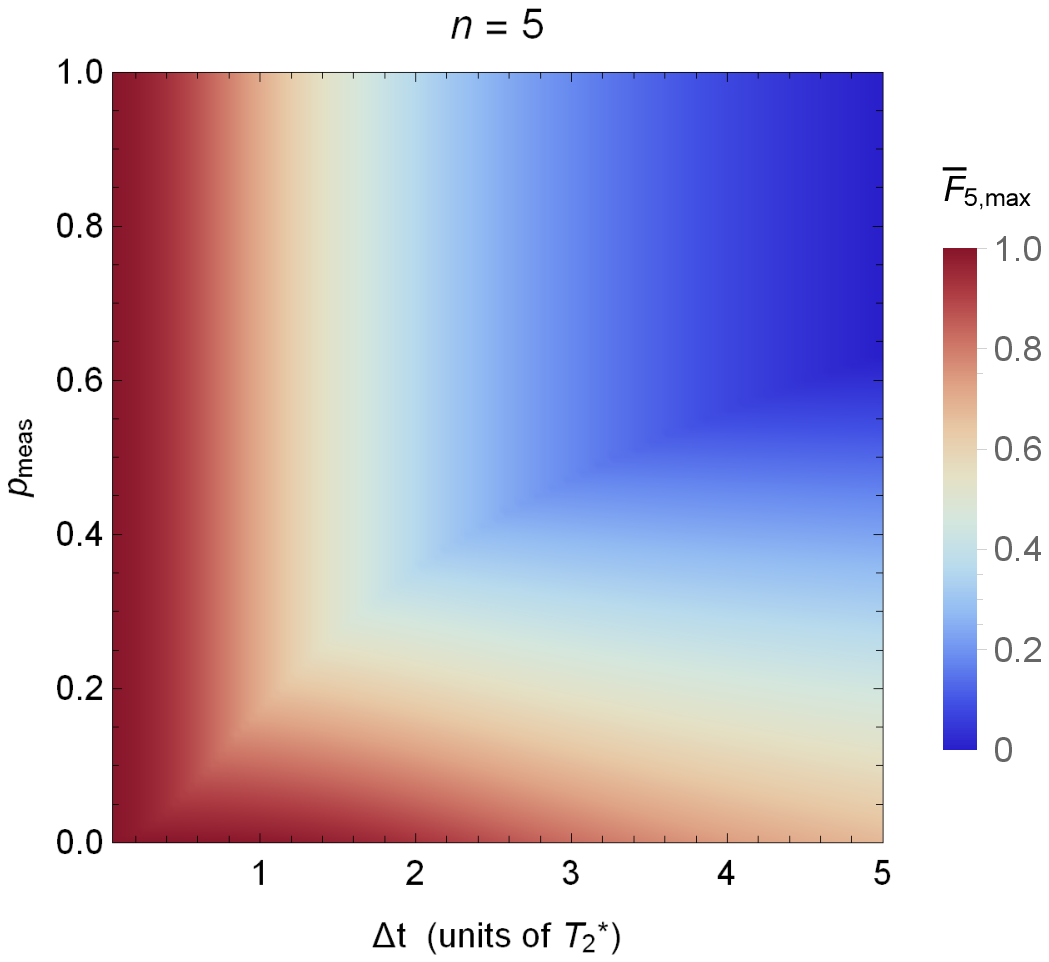}
    }
    \hfill
    \subfloat{
    \includegraphics[height=0.4\textwidth]{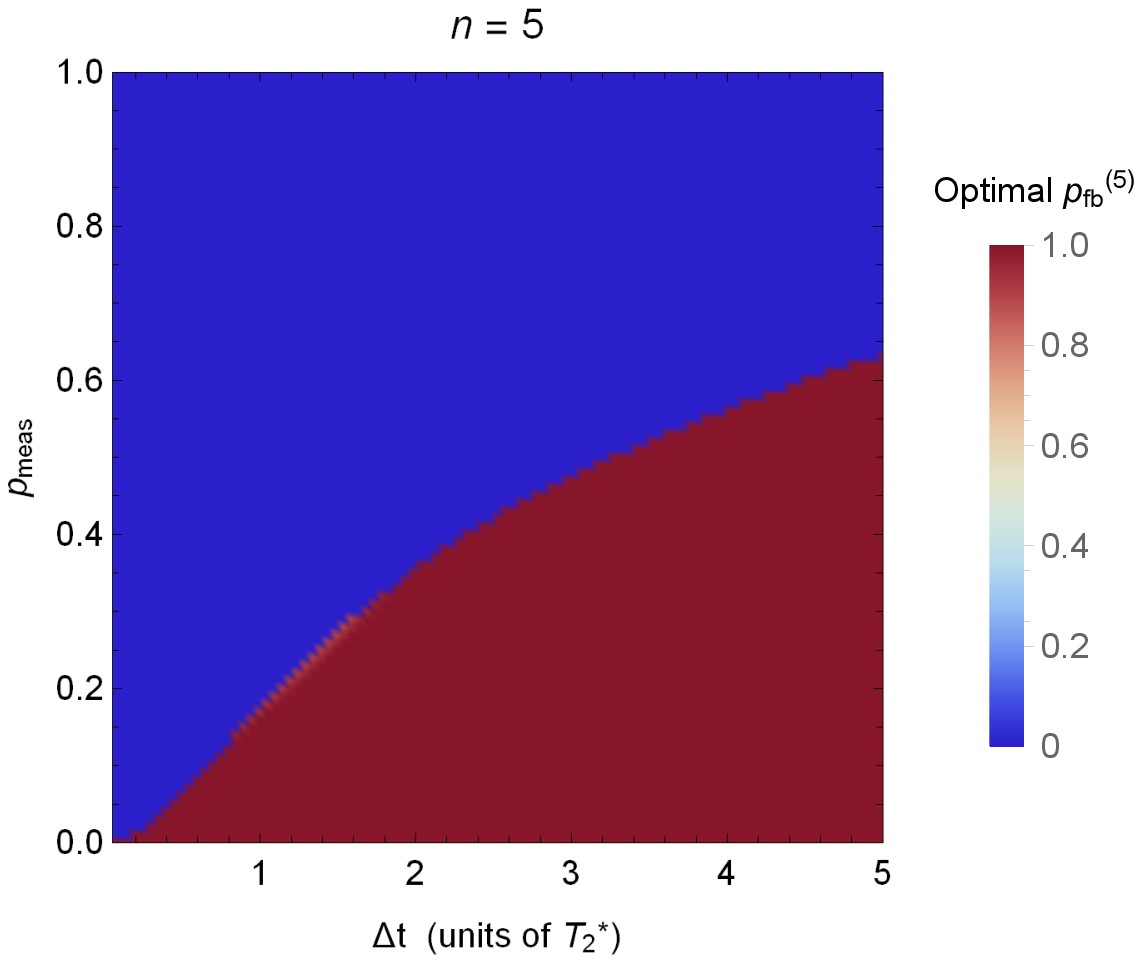}
    }
\end{figure*}
\begin{figure*}
    \subfloat{
    \includegraphics[height=0.4\textwidth]{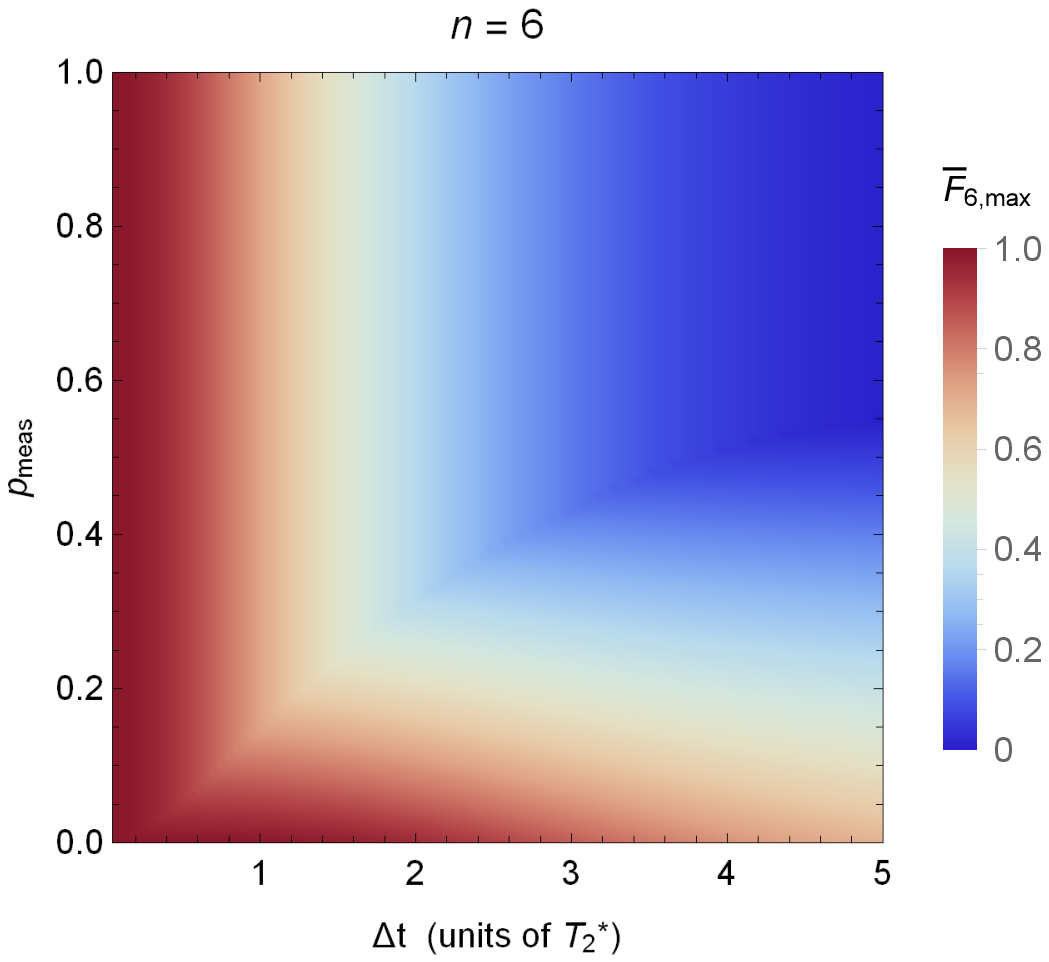}
    }
    \hfill
    \subfloat{
    \includegraphics[height=0.4\textwidth]{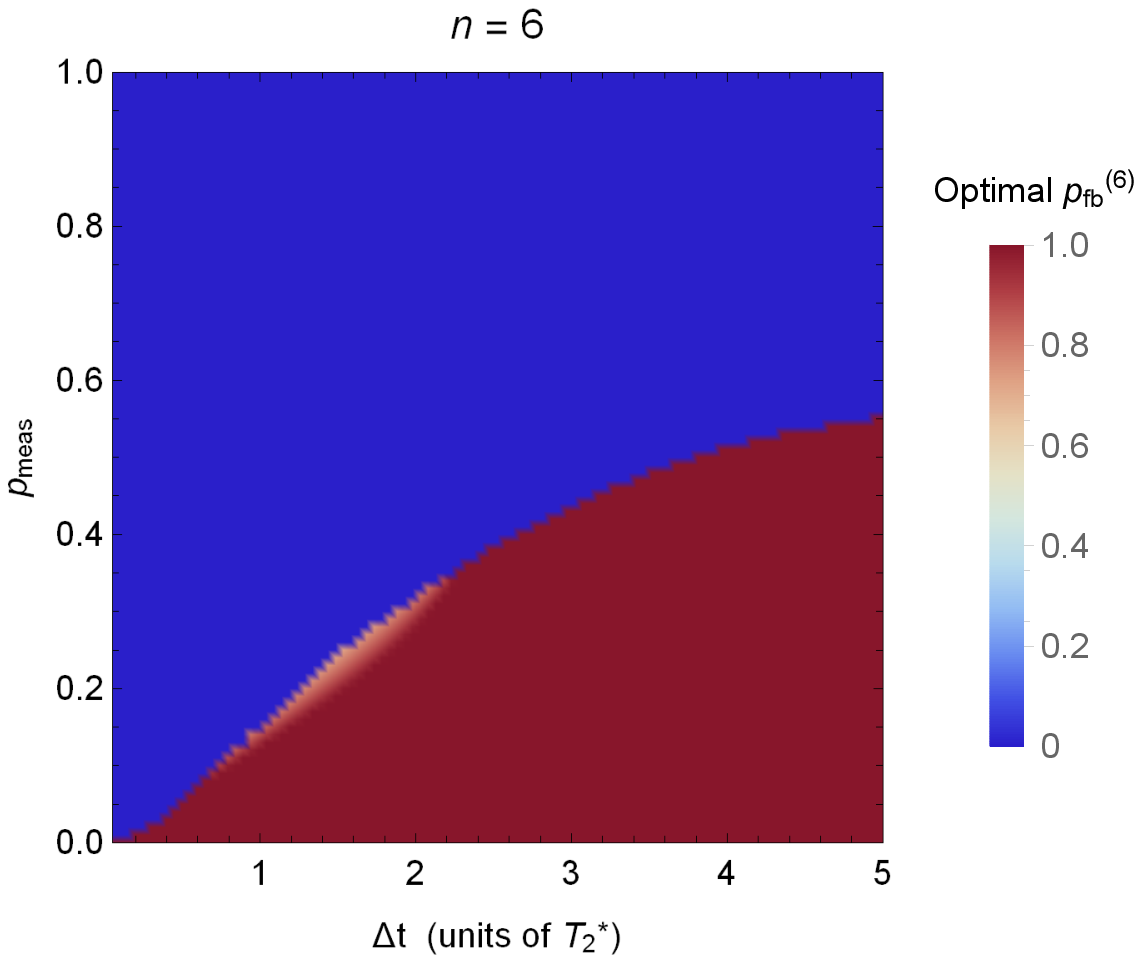}
    }
\end{figure*}
\begin{figure*}
    \subfloat{
    \includegraphics[height=0.4\textwidth]{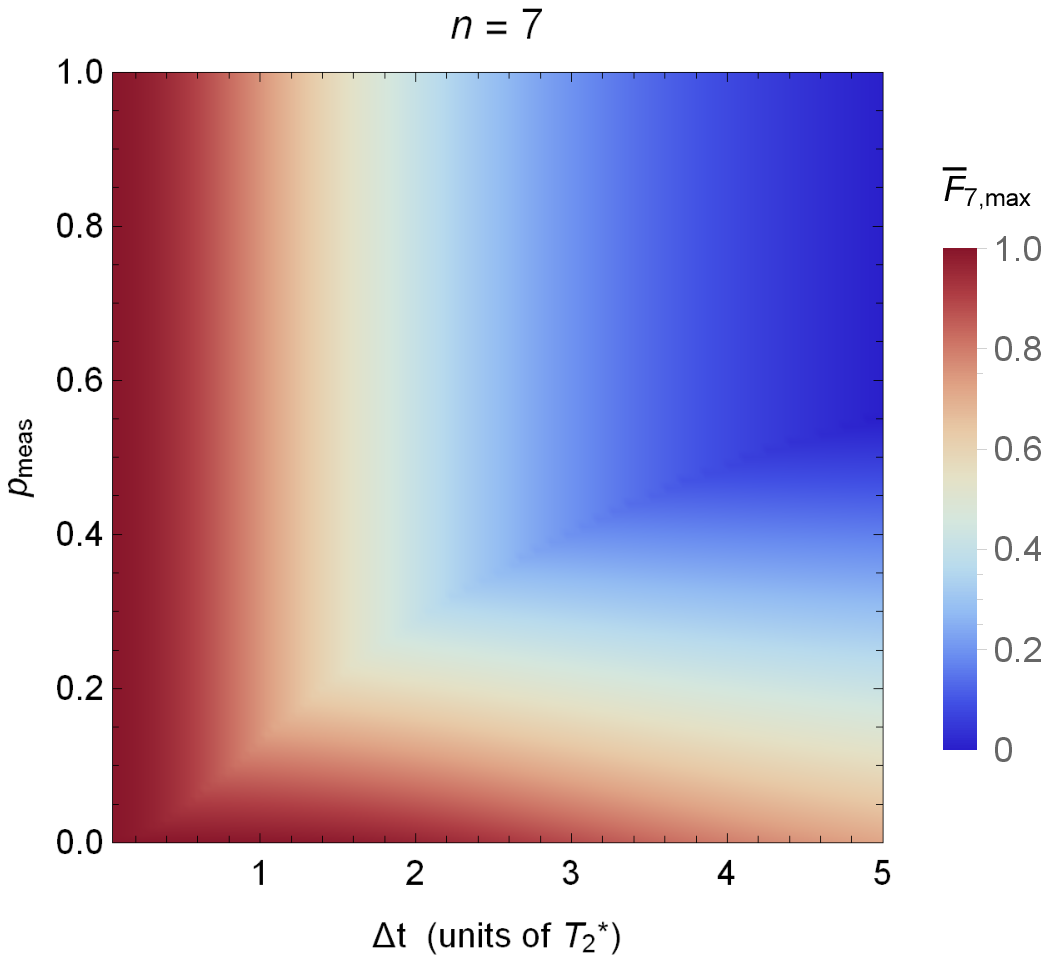}
    }
    \hfill
    \subfloat{
    \includegraphics[height=0.4\textwidth]{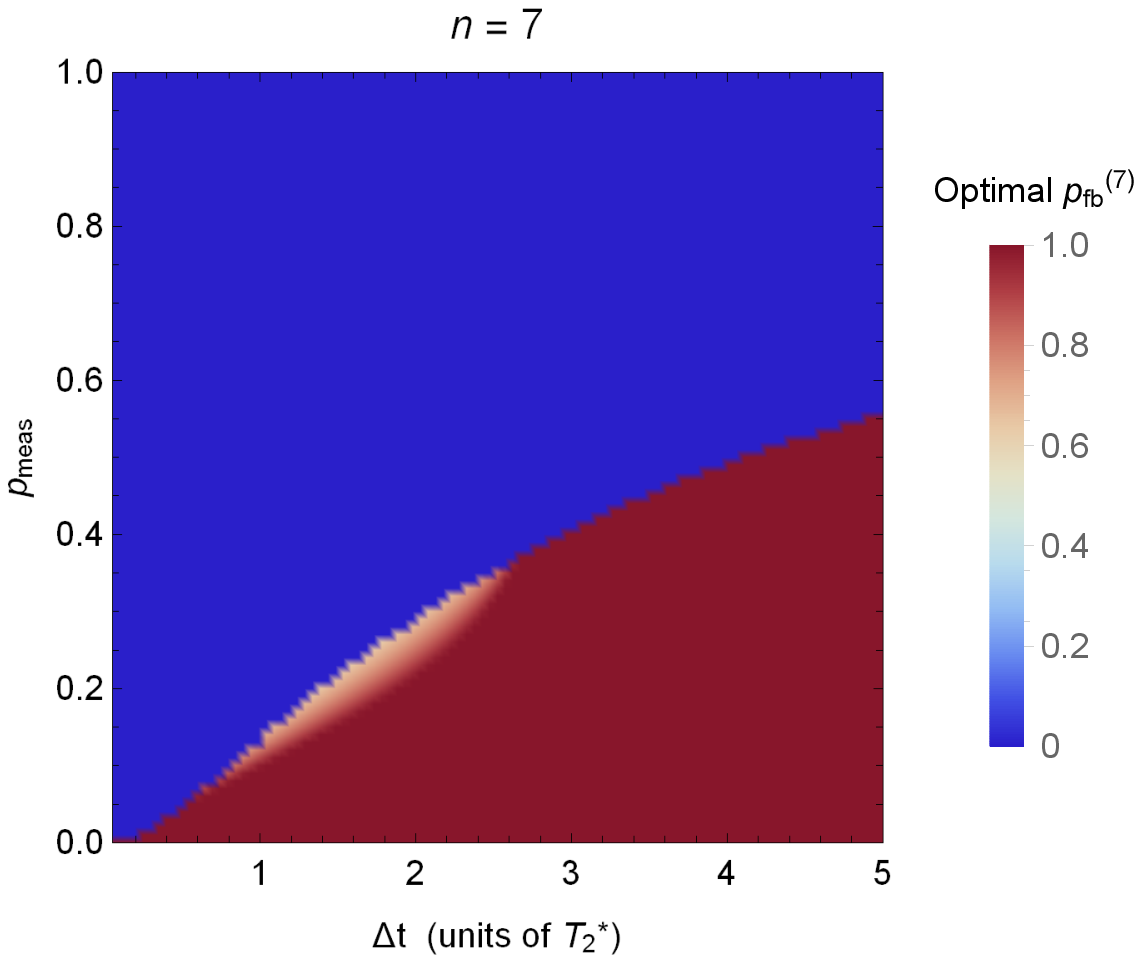}
    }
\end{figure*}
\begin{figure*}
    \subfloat{
    \includegraphics[height=0.4\textwidth]{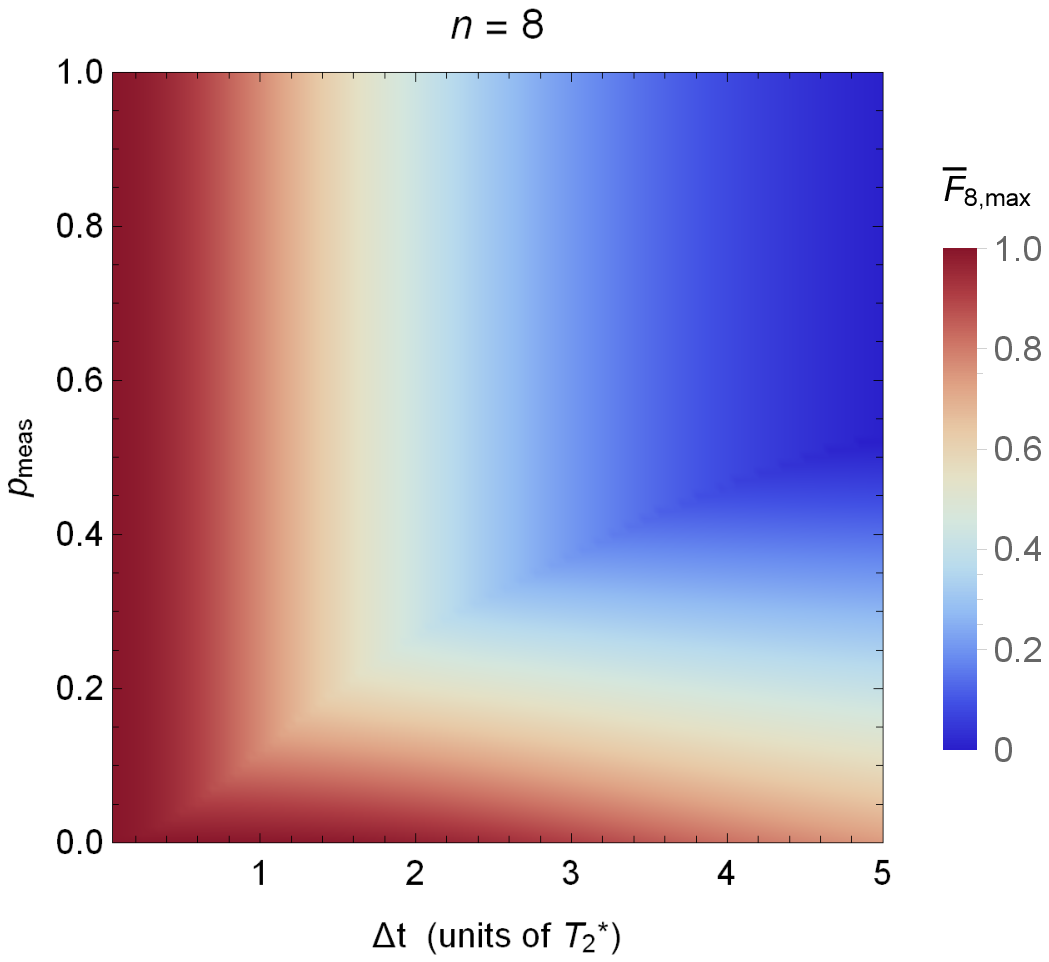}
    }
    \hfill
    \subfloat{
    \includegraphics[height=0.4\textwidth]{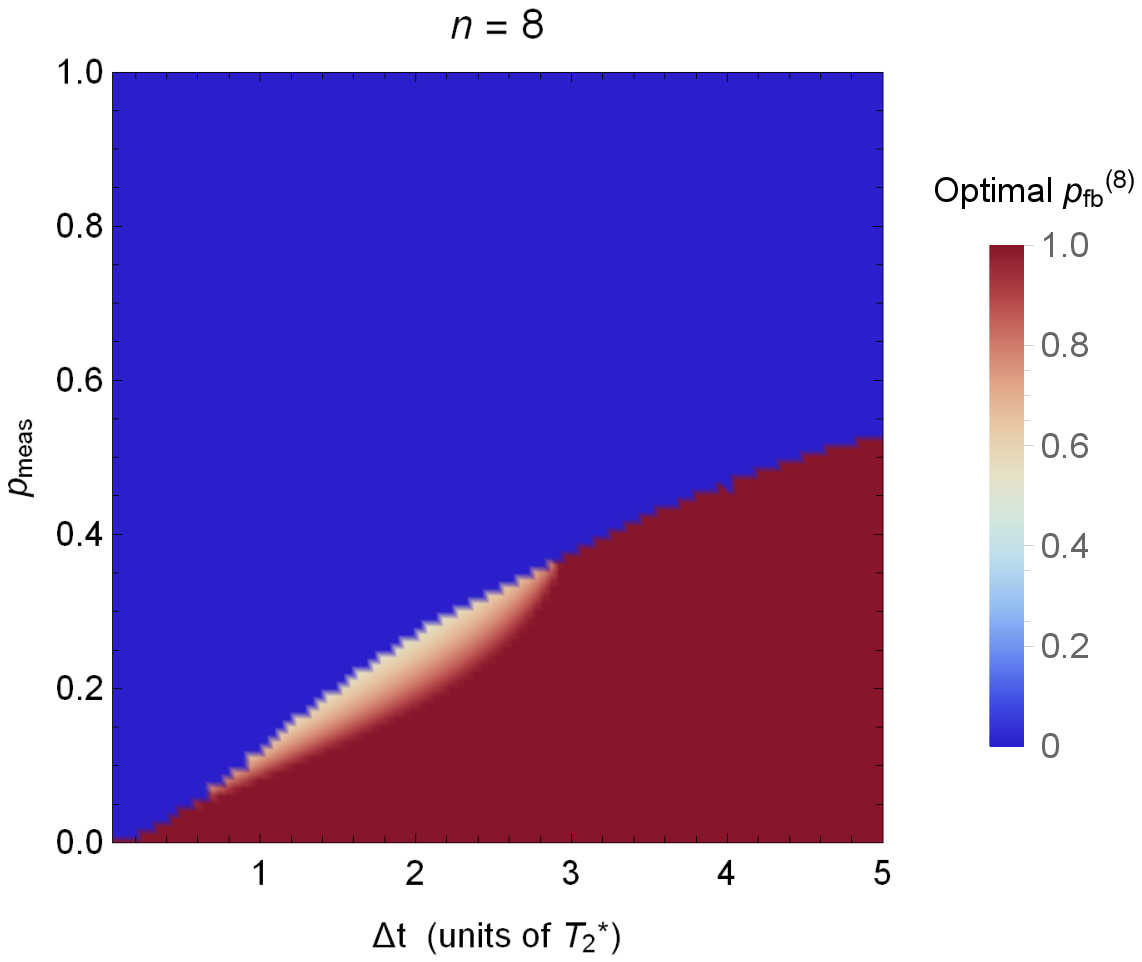}
    }
\end{figure*}
\begin{figure*}
    \subfloat{
    \includegraphics[height=0.4\textwidth]{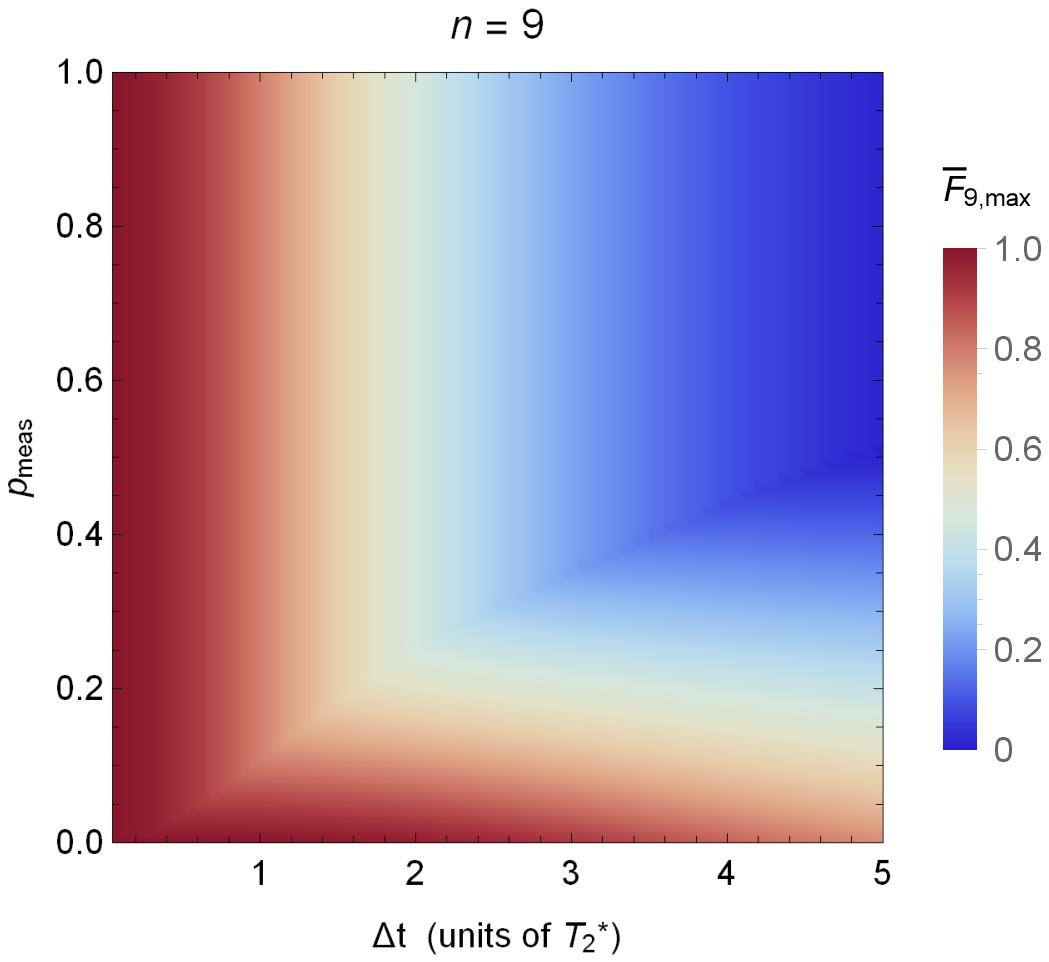}
    }
    \hfill
    \subfloat{
    \includegraphics[height=0.4\textwidth]{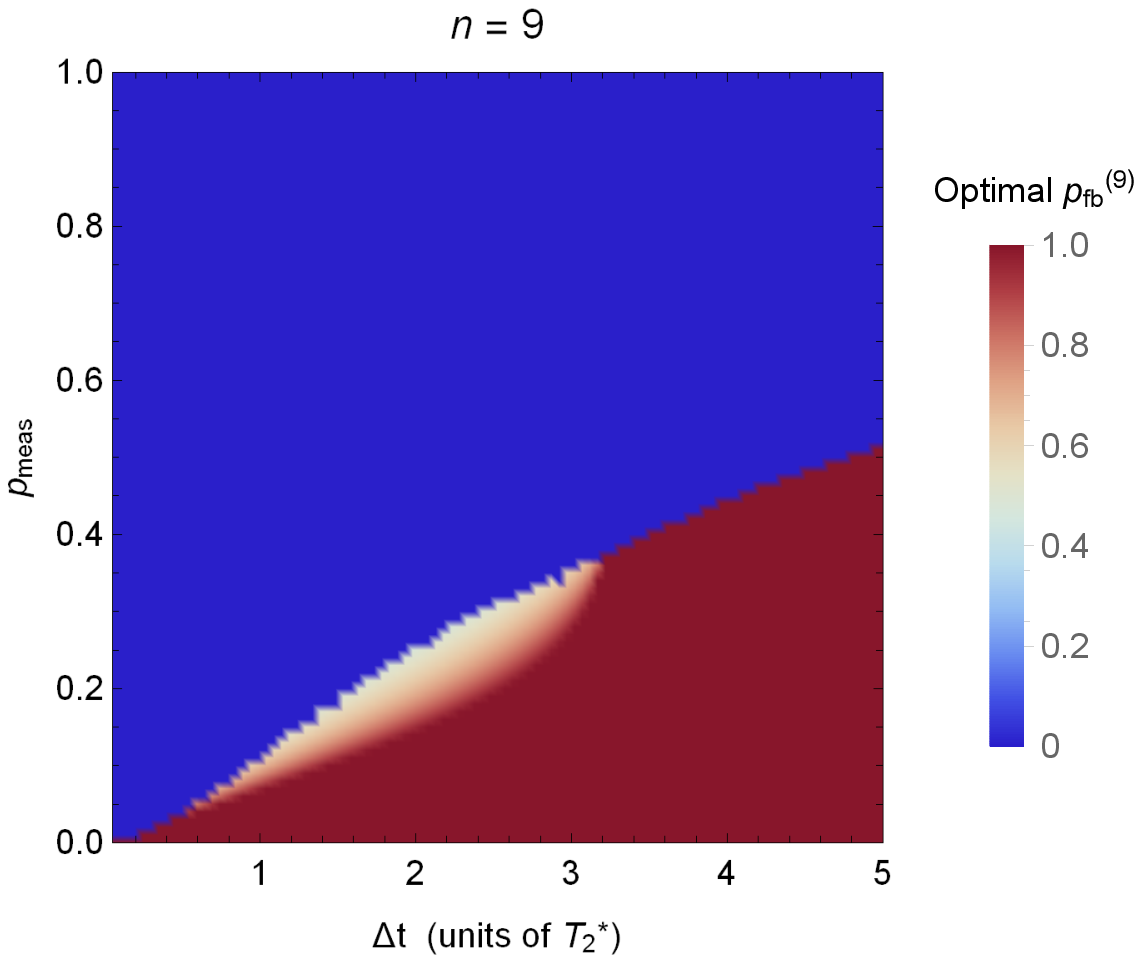}
    }
\end{figure*}

\end{document}